\documentclass[fp,twocolumn,letter]{jpsj3}
\usepackage{color}
\usepackage{bm}
\usepackage[version = 4]{mhchem}
\usepackage[resetlabels,labeled]{multibib}

\title{Large Reduction in the $a$-axis Knight Shift on UTe$_2$ with $T_{\rm c}$ = 2.1~K}
\author{ Hiroki Matsumura$^{1}$\thanks{matsumura.hiroki.75r@st.kyoto-u.ac.jp}, Hiroki Fujibayashi$^1$\thanks{fujibayashi.hiroki.w49@kyoto-u.jp}, Katsuki Kinjo$^{1}$\thanks{Present adress: Institute of Multidisciplinary Research for Advanced Materials, Tohoku University, Sendai, Miyagi 980-8577, Japan}, Shunsaku Kitagawa$^{1}$, Kenji Ishida$^1$\thanks{kishida@scphys.kyoto-u.ac.jp}, \\Yo Tokunaga$^{2}$, Hironori Sakai$^{2}$, Shinsaku Kambe$^{2}$, Ai Nakamura$^{3}$, Yusei Shimizu$^{3}$, \\Yoshiya Homma$^{3}$, Dexin Li$^{3}$, Fuminori Honda$^{3,4}$, and Dai Aoki$^{3,5}$}
\inst{
$^1$Department of Physics, Graduate School of Science, Kyoto University, Kyoto 606-8502, Japan \\
$^2$Advanced Science Research Center, Japan Atomic Energy Agency, Tokai, Ibaraki 319-1195, Japan \\
$^3$Institute for Materials Research, Tohoku University, Oarai, Ibaraki 311-1313, Japan\\
$^4$Central Institute of Radioisotope Science and Safety, Kyushu University, Fukuoka 819-0395, Japan\\
$^5$Universit\'e Grenoble Alpes, CEA, IRIG, PHELIQS, F-38000 Grenoble, France} 
\date{\today}
\abst{
Spin susceptibility in the superconducting (SC) state was measured in the higher-quality sample of uranium-based superconductor UTe$_2$ by using Knight-shift measurements for a magnetic field $H$ along all three crystalline axes.
In the higher-quality sample, the SC transition temperature $T_{\rm c}$ is about 2.1~K, and the residual electronic term in the specific heat is almost zero. 
The NMR linewidth becomes narrower and is almost half of that in the previous sample with $T_{\rm c} \sim 1.6$~K when $H \parallel a$ and $c$.
Although the Knight-shift behavior was not so different from the previous results for $H \parallel b$, and $c$, a large reduction in Knight shift along the $a$ axis was observed, in contrast with the previous $a$-axis Knight shift result.  
We discuss the origin of the difference between the previous and present results, and the possible SC state derived from the present results.
}

\begin{document}
\maketitle

Superconductivity in UTe$_2$ was discovered at the end of 2018\cite{RanScience2019}. 
UTe$_2$ crystallizes in the orthorhombic, centrosymmetric structure (space group \#71, $Immm$), and U atoms form parallel linear chains along the [100] $a$ axis [Fig.~1(a)].
After the discovery, UTe$_2$ is considered to be a spin-triplet superconductor, because its physical and superconducting (SC) properties are similar to those in the U-based ferromagnetic (FM) superconductors\cite{AokiJPSJ2019Rev}.
In fact, the experimental results of multiple SC phases\cite{BraithwaiteCommPhy2019,ThomasSciAdv2020}, spontaneous time-reversal symmetry breaking\cite{HayesScience2021}, and the chiral Majorana edge and surface state\cite{JiaoNature2020} suggest spin-triplet superconductivity with spin and/or orbital degrees of freedom.

We have measured the $^{125}$Te-NMR Knight shifts of UTe$_2$ to investigate the spin susceptibility in the SC state\cite{NakamineJPSJ2019, NakaminePRB2021, NakamineJPSJ2021}.
The Knight shift probing the static field at the nuclear site is one of the most reliable measurements of spin susceptibility in the SC state.
In a single crystalline sample with the SC transition temperature $T_{\rm c} \sim 1.6$~K, we have reported a slight decrease in the Knight shift for $^{125}$Te NMR along the $b$ and $c$ axes ($K_b$ and $K_c$, respectively) at a low magnetic field of $\mu_0 H$ = 1 T in the SC state\cite{NakamineJPSJ2019, NakaminePRB2021}.
We showed that the decrease in $K_b$ and $K_c$ is much smaller than those expected in spin-singlet superconductors, suggesting that UTe$_2$ is a spin-triplet superconductor with the finite components of $\hat{b}$ and $\hat{c}$ in the $\bm{d}$ vector.
Here, $\bm{d}$ vector is the SC order parameter of the spin-triplet pairing. 
In addition, we found that $K_a$ almost follows the normal-state temperature dependence, suggesting that the main spin component of the SC pairing is along the $a$ axis. 

However, it was found that an {\it early-stage} sample with $T_{\rm c} \sim 1.6$~K, in which nearly half of the electronic term in the specific heat remains at $T \rightarrow 0$, includes a non-negligible U deficiency from the careful X-ray diffraction (XRD) measurements on various quality samples on UTe$_2$\cite{Haga2022JPCM}.
After many efforts, the high-quality sample was successfully prepared quite recently.
In the high-quality sample, $T_{\rm c}$ is $= 2.1$~K and the residual electronic term in the specific heat is almost zero\cite{Haga2022JPCM,Sakai2022PRM}.
This is expected to be a disorder-free sample from the relationship between $T_{\rm c}$ and the residual electronic term\cite{aoki2021JPCMrev}.
Therefore, it is essentially important to investigate the Knight-shift behavior in the disorder-free sample to conclude what type of the SC pairing state is realized on UTe$_2$.  
 
\begin{figure}[tbp]
\begin{center}
\includegraphics[width=8cm]{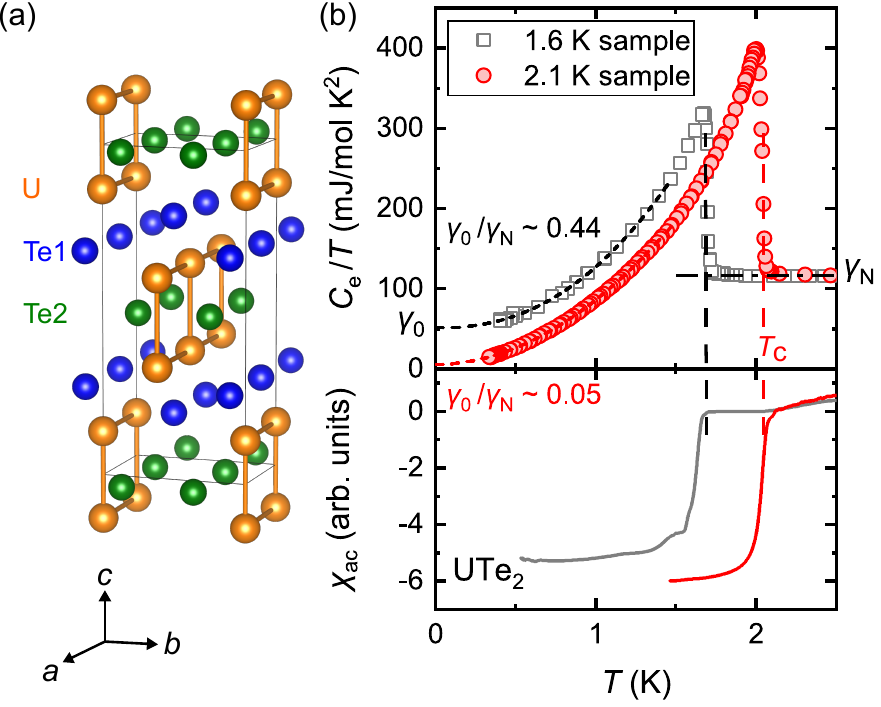}
\end{center}
\caption{(Color online) (a) Crystal structure of UTe$_2$.
(b) Temperature dependence of the electronic term of the specific heat divided by temperature $C_{\rm e} / T$ and the ac susceptibility measured with variation in resonance frequency of NMR tank circuits in the 1.6~K and 2.1~K samples. }
\label{f1}
\end{figure}

$^{125}$Te-enriched high-quality samples were prepared with the newly developed molten salt flux method\cite{Sakai2022PRM}.  
Natural U and 99.9~\% $^{125}$Te-enriched metals were used as starting materials for the present sample. 
The single crystals under an optimized growth condition with excess uranium exhibit an SC transition at $T_{\rm c}$ = 2.1~K which was determined with the specific heat and ac susceptibility measurements as shown in Fig.~\ref{f1}(b). 
This is the highest $T_{\rm c}$ reported in UTe$_2$\cite{aoki2021JPCMrev}. 
This sample shows a quite small residual electronic term well below $T_{\rm c}$, indicating that the increase in $T_{\rm c}$ is ascribed to the reduction of the disorder due to the uranium deficiency.
The detailed analysis of the specific-heat result in the 2.1~K sample is described in the supplemental materials\cite{SM1}.

The $^{125}$Te (nuclear spin $I$ = 1/2, gyromagnetic ratio $^{125}\gamma/2\pi$ = 13.454 MHz/T)-NMR measurements were performed on two single crystals of size $2 \times 1.2 \times 1$ mm$^3$ and $3 \times 1 \times 0.5$ mm$^3$.
We reported that two $^{125}$Te-NMR signals were observed in UTe$_2$ because of the presence of the two inequivalent crystallographic Te sites, as seen in Fig.~\ref{f1}(a).
Note that the peak assignment [Te(I) and Te(II)] does not correspond to the crystallographical site (Te1 and Te2).
Although we measured both signals, there were no qualitative differences in the Knight-shift results.
Thus, we focused on the Knight-shift results in the $^{125}$Te(II)-NMR signal for ensuring the accuracy of the data, as the Knight shift and NMR-signal intensity at the Te(II) are larger than those at the Te(I).   
The NMR spectra as a function of frequency were obtained using the Fourier transform of a spin-echo signal observed after a radio-frequency (RF) pulse sequence at a fixed magnetic field. 
The magnetic field was calibrated using a $^{65}$Cu ($^{65}\gamma /2\pi = 12.089$ MHz/T)-NMR signal from the NMR coil. 
The NMR spectra in the SC state were recorded in a field-cooling process.
The sample was rotated in the $ab$ and $bc$ planes to apply the magnetic field $H$ precisely along each axis using a split-pair magnet with a single-axis rotator, and the Knight shift was determined by the peak position of the NMR spectrum.
For reliable NMR measurements in the SC state, the energy of the RF pulses was reduced to ensure that the NMR results were unchanged by the power of the RF pulses. 
Furthermore, we experimentally confirmed the superconductivity just after the NMR RF pulses using a technique reported in previous studies\cite{IshidaJPSJ2020,NakamineJPSJ2019}. 

\begin{figure}[tbp]
\begin{center}
\includegraphics[width=8cm]{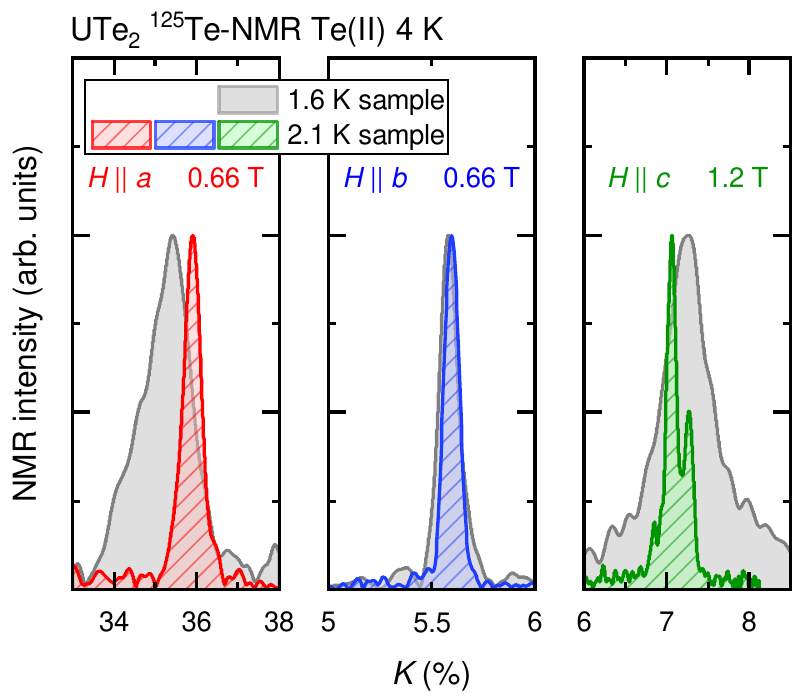}
\end{center}
\caption{(Color online) $^{125}$Te-NMR spectrum with the larger Knight shift [Te(II)] measured in the 1.6~K and 2.1~K samples when $H$ is applied to three crystalline axes. In $H \parallel c$, the 1.6~K sample shows the broad spectrum due to the overlapping of Te(I) and Te(II)-NMR peaks, but the two peaks were separated in the 2.1~K sample. All spectra were measured at 4~K. }
\label{f2}
\end{figure}

Figure \ref{f2} shows the Te(II)-NMR spectra in $H$ parallel to all three crystalline axes measured in the previous {\it early-stage} and present {\it high-quality} samples. 
In this paper, we refer to the previous and present samples as the 1.6~K and 2.1~K samples, respectively, after the $T_{\rm c}$ of the samples.
The Knight-shift values are not so different between the two samples, indicating that the density of states and electron correlations are almost the same.
This is consistent with the specific-heat result in the normal state\cite{aoki2021JPCMrev}. 
On the other hand, the NMR linewidth is different between the two samples.
As shown in Fig.~\ref{f2}, the NMR spectra of the 1.6~K sample are broad and asymmetric, but those of the 2.1~K sample are narrow and symmetric. 
Particularly, a clear two-peak structure was recognized in the NMR spectrum of the 2.1~K sample in $H \parallel c$.
The two peaks in the 2.1~K sample arise from the Te(II) and Te(I) signals, and the two peaks become broader and overlap each other in the 1.6~K sample.
The angle dependence of the two peaks, which was used for the signal assignment, is shown in supplemental materials\cite{SM2}.
The sharp NMR spectrum improved the accuracy of Knight-shift measurements significantly.   
As mentioned above, the presence of a small amount (order of 1\%) of U deficiency was concluded from the XRD measurements in the 1.6~K sample\cite{Sakai2022PRM}, and thus, it is reasonable to consider that the broad linewidth in the 1.6~K sample is ascribed to the U deficiency.
As the $H \parallel a$ $(b)$ spectrum of the 1.6~K sample is asymmetric with the tail to the smaller (larger) $K$ direction, it is considered that the less U-deficiency part of the sample possesses more anisotropic spin susceptibility than the U-deficiency part, suggesting that the anisotropic spin susceptibility (larger $K_a$ and smaller $K_b$) is important for the occurrence of the superconductivity in UTe$_2$.

\begin{figure*}[t]
\begin{center}
\includegraphics[width=15cm]{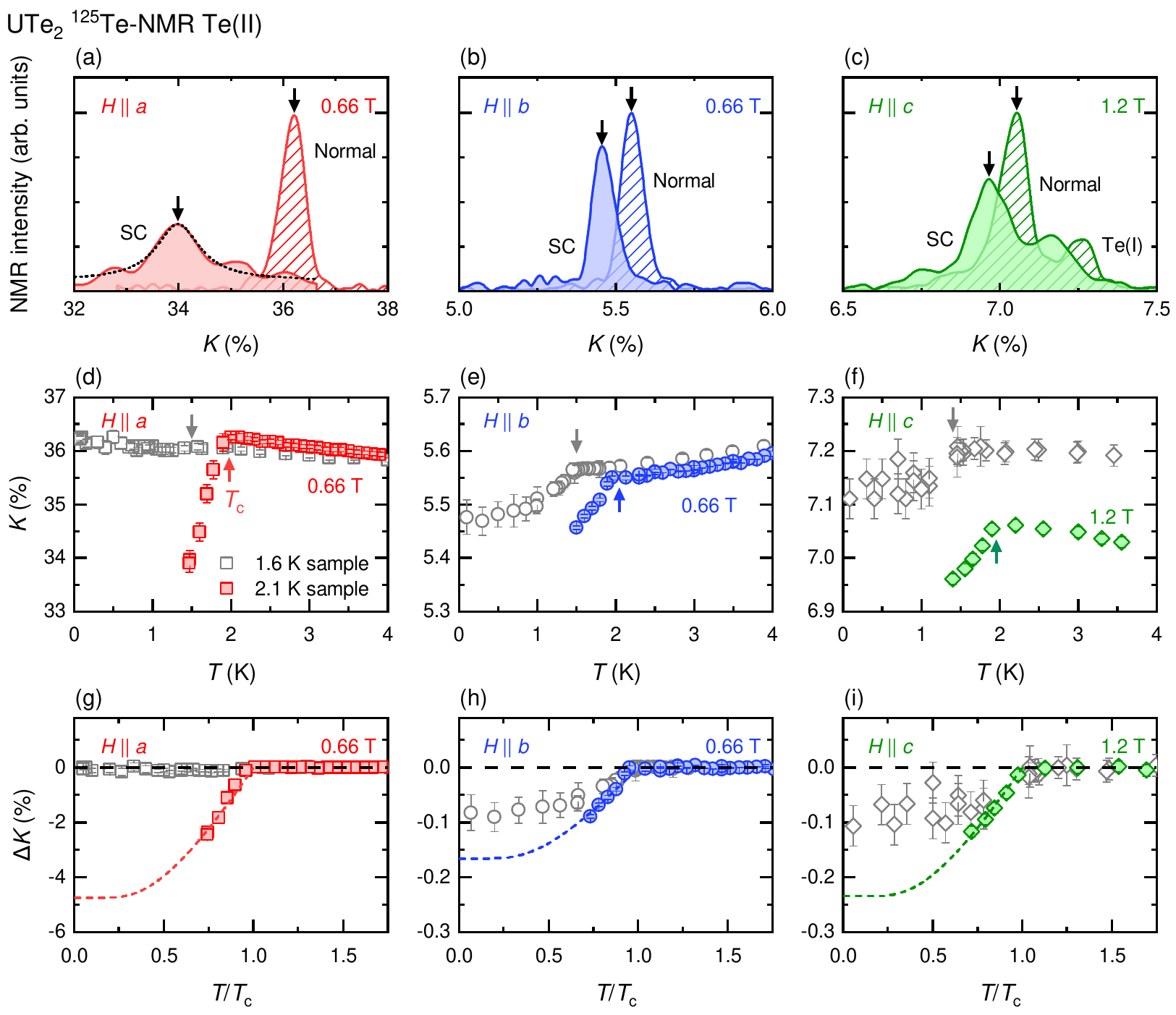}
\end{center}
\caption{(Color online) (a)-(c) $^{125}$Te-NMR spectrum measured at 2.3~K ($> T_{\rm c}$, normal state) and 1.5~K ($ < T_{\rm c}$, SC state) in the 2.1~K sample in $H$ applied parallel to three crystalline axes.
The intensity of the spectrum was normalized by signal area.
The arrows show the peaks to determine the Knight-shift values.
The dotted curve in (a) is a Gaussian fitting of the peak.
(d)-(f) Temperature dependence of the Knight shift in $H \parallel$ $a$, $b$, and $c$. The results of the 1.6~K sample are shown with the gray-color symbols. The arrows show $T_{\rm c}$ determined with the AC susceptibility measurements. (g) - (i) The Knight-shift variation ascribed to the SC transition ($\Delta K$) against the normalized $T$ by $T_{\rm c}$. In the figures, $T$ dependence of the normal-state Knight shift [$K_{\rm N}$($T$)] is subtracted from the above observed $K$($T$) [$\Delta K(T) \equiv K(T) -K_{\rm N}$($T$)]. The expected reduction at $T = 0$ was evaluated from the extrapolation with the Yosida function shown by the broken curves, and the broken lines are $\Delta K = 0$.}
\label{f3}
\end{figure*}

The NMR spectra in the 2.1~K sample measured at 2.3~K ($ > T_{\rm c}$) and 1.5~K ($ < T_{\rm c}$) in $H$ parallel to the three crystalline axes are shown in Figs.~\ref{f3}(a)-(c).
A clear spectrum shift below $T_{\rm c}$ was observed in all three directions.
Figures \ref{f3} (d)-(f) show the $T$ dependence of the Knight shift along three crystalline axes $K_i$ ($i$ = $a$, $b$, and $c$) in the 2.1~K samples, and the previous results in the 1.6~K sample are also plotted in the figures\cite{NakamineJPSJ2021,NakaminePRB2021,FujibayashiUTe2}.
The apparent difference in $K_{c}$ between the 1.6~K and 2.1~K samples, even in the normal state is mainly due to how the Knight shift is determined: the Knight shift for the 1.6~K sample was determined by the broad peak arising from the overlap of the Te(II) and Te(I) peaks, while the Knight shift for the 2.1~K sample was determined by the sharp peak of the Te(II) signal. 
Thanks to the increase in $T_{\rm c}$, the Knight shift starts to decrease from the higher temperature and the reduction becomes more significant in the 2.1~K sample. 
To estimate the variation of the Knight shift ascribed to the SC transition [$\Delta K_i$($T$)] ($i = a, b$ and $c$), $T$ dependence of the normal-state Knight shift [$K_{{\rm N}, i}$($T$)] is subtracted from the observed $K_i$($T$) [$\Delta K_i(T) \equiv K_i(T) -K_{{\rm N}, i}$$(T)$].

Figures \ref{f3}(g)-(i) show the plot of the $\Delta K_i$ against the $T$ normalized to $T_{\rm c}$ in the 1.6 and 2.1~K samples.  
Although the decrease in $\Delta K_b$ and $\Delta K_c$ was observed in both samples, the $K_a$ behavior was quite different between the two samples: $K_a$ decreases with $\Delta K_a \sim 2$\% at 1.5~K in the 2.1~K sample, but $K_a$ was unchanged in the 1.6~K sample. 
The reduction in $K_a$ of the 2.1~K sample is so sudden that we missed the $^{125}$Te-NMR in the SC state at first. 
By measuring the $T$ variation of the Knight shift by changing the angle from the $b$ to $a$ axis in the $ab$ plane, we succeeded in following the $^{125}$Te-NMR spectrum in the SC state and in measuring the $T$ variation of $K_a$.  

Here, we discuss the origin of the different behavior between the two samples.
As previously reported [Fig.~\ref{f3}(d)]\cite{FujibayashiUTe2}, $K_a$ in the 1.6~K sample continued to increase  down to 75 mK with a small kink at $T_{\rm c}$, which originates from the SC diamagnetic effect. 
Although we believe that the continuous increase in the SC state is an intrinsic behavior in UTe$_2$ after we measured $K_a$ in the 2.1~K sample, we should consider the following two possibilities: one is that we observed the NMR signal arising from the non-SC part induced by the U deficiency and missed the NMR signal from the SC part in the 1.6~K sample.

This possibility seems to be consistent with the results suggested from the comparison of the electronic term in specific heat ($C_{\rm e}$) between the two samples.
As seen in Fig.~\ref{f1}(b), the $C_{\rm e}/T$ in the 2.1~K sample shows the larger jump at $T_{\rm c}$ and almost no residual $\gamma$ term at $T \rightarrow 0$.
In addition, it is reasonable to consider that $C_{\rm e}/T$ in the 2.1~K sample is the ideal behavior in UTe$_2$ from the satisfaction of the entropy balance below $T_{\rm c}$, as discussed in the supplemental materials\cite{SM1}.
Assuming that the 1.6~K sample consists of the SC and non-SC parts and that the jump height of $C_{\rm e}/T$ in the SC part is independent of sample quality, the non-SC fraction is estimated to be 29\% from the jump of $C_{\rm e}/T$ at $T_{\rm c}$, which is not so far from the value of residual $\gamma_0$ divided by the normal-state $\gamma_{\rm N}$ ($\gamma_0/\gamma_{\rm N} \sim$ 0.44)\cite{Cairns2020JPCM}.
We might follow the $^{125}$Te-NMR signal arising from the non-SC part in the sample, since the intensity of the NMR signal from the SC part is much reduced by the diamagnetic effect in general.
The unusual decrease of the NMR-spectrum linewidth below $T_{\rm c}$ in the 1.6 K sample\cite{FujibayashiUTe2} might be interpreted with this possibility.
The other possibility is that the spins of the SC pairs are easily aligned to the applied field, as $\mbox{\boldmath{$d$}}$-vector is considered not to be strongly fixed.   
We discuss this possibility later.  
To identify the origin of the different behavior between two samples, it is crucially important to investigate the dependence of $K_a$ against the U-deficiency and/or $H$ dependence of $K_a$. 

\begin{table}[htb]
\begin{center}
\caption{\label{t1}Classification of the odd-parity SC order parameters for point group with $D_{\rm 2h}$\cite{IshizukaPRL2019}. The irreducible representation (IR) and its basis functions are listed. To clarify the dominant SC spin component, a spin component perpendicular to the $\mbox{\boldmath$d$}$ vectors is also shown.}
\vspace{5mm}
  \begin{tabular}{cccc}\hline \hline 
 \multicolumn{1}{c}{$D_{\rm 2h}$ (zero field)} \\ \hline 
IR & Basis functions & & SC spin comp.\\ \hline
$A_{\rm u}$ & $k_a \hat{a}$, $k_b \hat{b}$, $k_c \hat{c}$ & & \\
$B_{\rm 1u}$ & $k_b \hat{a}$, $k_a \hat{b}$ & & $c$ \\
$B_{\rm 2u}$ & $k_a \hat{c}$, $k_c \hat{a}$ & & $b$ \\
$B_{\rm 3u}$ & $k_c \hat{b}$, $k_b \hat{c}$ & & $a$ \\ \hline \hline
\\ \\
  \end{tabular}
  \end{center}
\end{table}

Now we discuss the implication of the present results.
The possible spin-triplet SC symmetries based on irreducible representation in UTe$_2$ with a $D_{\rm 2h}$ point group are listed in Table \ref{t1} \cite{IshizukaPRL2019}.  
In the previous paper\cite{FujibayashiUTe2}, we suggested that the spin-triplet $B_{\rm 3u}$ state is the most promising state, in which $\bm{d}$ vector has the $\hat{b}$ and $\hat{c}$ components.
This conclusion should be modified by the present Knight-shift results in the 2.1~K sample.
The decrease fraction ascribed to superconductivity in the spin susceptibility against the normal-state susceptibility $\Delta \chi_i / \chi_{{\rm N},i}$ is estimated from the relation of $\Delta K_i / K_{{\rm N},i} = \Delta \chi_i / \chi_{{\rm N}, i}$ = (0.13, 0.020, 0.028). 
Here, $\Delta K_i$ is estimated from the extrapolation of the Knight-shift reduction at $T \rightarrow 0$ by assuming the Yosida function with conventional BCS model parameters \cite{YosidaPhysRev1957} as shown in Figs.~\ref{f3}(g)-(i), and the SC diamagnetic effect was taken into account as discussed in supplemental materials\cite{SM3}.
The spin susceptibility decreases in all crystalline directions, suggesting the spin-singlet pairing or the spin-triplet $A_{\rm u}$ state, in which $\bm{d}$ vector has all crystalline-axis components.
However, the possibility of the spin-singlet pairing state can be excluded for the following reasons.
First, the coherence peak of $1/T_1T$ just below $T_{\rm c}$ is absent in the 2.1~K sample as in the 1.6~K sample\cite{NakamineJPSJ2019}, excluding the conventional $s$-wave superconductivity\cite{SM4}.    
In addition, the spin-singlet pairing should be destroyed by the Pauli-depairing effect and the Pauli-depairing field $H_{\rm P}$ is roughly estimated from the relations of $H_{\rm P} = H_{\rm c}/\sqrt{\Delta \chi}$ with the thermodynamic SC critical field $H_{\rm c}$ and the reduction of the spin susceptibility in the SC state $\Delta \chi$. 
When we adopt $\mu_0 H_{\rm c} \sim 76.8$ mT derived from the specific-heat result and $\Delta \chi$ along the $a$ axis, $\mu_0 H_{\rm P} \sim 1.9$ T is estimated.
This is much smaller than the upper critical field $\mu_0 H_{\rm c2} \sim 12$ T in $H \parallel a$. 
The absence of the Pauli-depairing effect at around $\mu_0 H_{\rm P} \sim 1.9$ T excludes the spin-singlet pairing, but is rather consistent with the spin-triplet $A_{\rm u}$ state with the spin degrees of freedom.
In the spin-triplet superconductivity, the spin component of the spin-triplet paring can be induced by the applied $H$, and thus, $\Delta K_a$ is immediately suppressed to avoid the Pauli-depairing effect when the interaction pinning the $\bm{d}$ vector to a certain direction is small. 
Such phenomenon, known as $\bm{d}$-vector rotation, was actually observed in the $H$ dependence of the Knight shift when $H \parallel b$\cite{NakamineJPSJ2021} and $c$\cite{NakaminePRB2021}. 

In addition, the present results are qualitatively consistent with the theoretical prediction.  
As discussed in the theoretical paper by Hiranuma and Fujimoto\cite{HiranumaJPSJ2021}, the present results of the large reduction in $K_a$ and the small reduction in $K_b$ and $K_c$ are consistent with the $A_{\rm u}$ state realized in the split bands by the strong spin-orbit coupling much larger than the SC gap.
This is because the spin susceptibility along the $a$-axis $\chi_{a}$, which consists of the intraband-contributions only, decreases significantly below $T_{\rm c}$, but $\chi_{b}$ and $\chi_{c}$, which consist of mainly interband contributions that are unchanged in the SC state, decreases slightly.

We comment on that the reduction in $K_a$ of the 2.1~K sample is unexpectedly larger than what we considered.
From the analogy of FM superconductors\cite{MineevPRB2002}, it is expected that $\hat{a}$ component of the $\bm{d}$ vector would be the smallest at least in zero field.
The present result of the large magnitude of $\Delta K_a / K_{{\rm N}, a}$ suggests that the $\hat{a}$ component of the $\bm{d}$ vector is the largest, which is opposite to the above expectation.
Alternatively, it can be interpreted that the large magnitude of $\Delta K_a /  K_{{\rm N}, a}$ might be a consequence of the largest spin component of the $a$ axis in the normal state. 
It is a crucially important to uncover the relation between the $\bm{d}$-vector direction and the spin anisotropy in the normal state.

Furthermore, $\Delta K_a$ is even larger than the simple estimation of the quasi-particle spin susceptibility $K_{\rm qp}$ based on the Fermi-liquid picture with experimental values.
As discussed in the previous papers\cite{Tou2005JPSJ,NakamineJPSJ2019,aoki2021JPCMrev}, $K_{\rm qp}$ that would be related to the superconductivity was estimated with the Sommerfeld coefficient $\gamma_{\rm n}$ in the normal state, the hyperfine-coupling constant $A_{\rm hf}$\cite{TokunagaJPSJ2019}.

$K_{\rm qp}$ along the $a$ axis is estimated as $\sim 1.33$ \% by assuming the effective moment $(\mu_{\rm eff})^2 = 3 \mu_{\rm B}^2$ for the free-electron value and the and the Wilson ratio $R = 1$ and by adopting $\gamma_{\rm n} = 116$ mJ/mol K$^2$ and $A_{\rm hf}$ = 4.7 T/$\mu_{\rm B}$ for the $a$ axis. \cite{TokunagaJPSJ2019}.
Compared to the estimated value of $K_{\rm qp}$, $\Delta K_a$ is more than 4 times larger than the above estimation of $K_{\rm qp}$.
This indicates that the $a$-axis spin component related to the superconductivity 
is enhanced by the Wilson Ratio. 
If large $K_{{\rm N}, a} = 36.3$ \% observed just at $T_{\rm c}$ is related to the superconductivity, $R \sim 27.3$ is obtained. 
Such an enhancement of $R$ is sufficiently conceivable when UTe$_2$ is considered as a magnetically enhanced metal, and $R \sim 30$ was actually reported in heavy-fermion YbRh$_2$(Si$_{0.95}$Ge$_{0.05}$)$_2$ near quantum criticality\cite{GegenwartPRL2005}.
As the finite spin susceptibility should remain at $T = 0$ in the $A_{\rm u}$ state of the spin-triplet pairing, it is important to estimate the residual spin susceptibility experimentally.
For a more accurate estimation of the spin component at $T = 0$, it is necessary to investigate $H$ dependence of $\Delta K_a$ at $H \rightarrow 0$ because of the possibility of the $\bm{d}$-vector rotation by small $H$.

In conclusion, the Knight shift in the SC state was measured in the higher-quality sample of UTe$_2$ with $T_{\rm c}$ = 2.1~K, and the unexpected large reduction was observed in $K_a$ below $T_{\rm c}$, which was in contrast to the previous result in the 1.6~K sample, while the Knight-shift behavior in $H \parallel b$, and $c$ was not so different between two samples.
The present results suggest that the possible SC state is spin-triplet $A_{\rm u}$ and that the invariance of the spin susceptibility in the SC state, which was obtained in the 1.6~K sample, originates from the non-SC fraction remaining in the sample, or the $\mbox{\boldmath{$d$}}$-vector rotation by small applied $H$.
The spin-triplet $A_{\rm u}$ state is the same pairing state as the superfluid $^3$He B-phase \cite{LeggettRMP1975}, and consistent with the recent full-gap behavior revealed by the thermal-conductivity measurements in the 2.1~K sample\cite{Suetsugu}. 
This SC state is a strong candidate of the topological superconductor\cite{IshizukaPRL2019}, in which the Majorana surface state is anticipated, and explains the observation of the surface state\cite{JiaoNature2020}. 

\section*{Acknowledgments}
The authors would like to thank J. Ishizuka, Y. Yanase, K. Machida, S. Fujimoto, Y. Kasahara, Y. Matsuda, V. P. Mineev, Y. Maeno, S. Yonezawa, J-P. Brison, G. Knebel, and J. Flouquet for their valuable inputs in our discussions. 
This work was supported by the Kyoto University LTM Center, Grants-in-Aid for Scientific Research (Grant No. JP19K03726, JP19H00646, JP20H00130, JP20KK0061, JP21K18600, JP22H01168, JP22H04933).
This work was also supported by JST SPRING (Grant Number JPMJSP2110).
H.~F. and K.~K. would like to acknowledge the support from the Motizuki Fund of Yukawa Memorial Foundation.

\section*{Authors Contribution}
H.~M. and H.~F. equally contributed to this work.

\renewcommand{\figurename}{Fig. S}
\setcounter{figure}{0}
\clearpage
\noindent
Supplemental materials of \\
{\bf ``Large Reduction in the $a$-axis Knight Shift on UTe$_2$ with $T_{\rm c}$ = 2.1~K''}

\section{Analyses of the specific heat result on the 2.1~K sample}
\begin{figure}[h]
\begin{center}
\includegraphics[width=7cm]{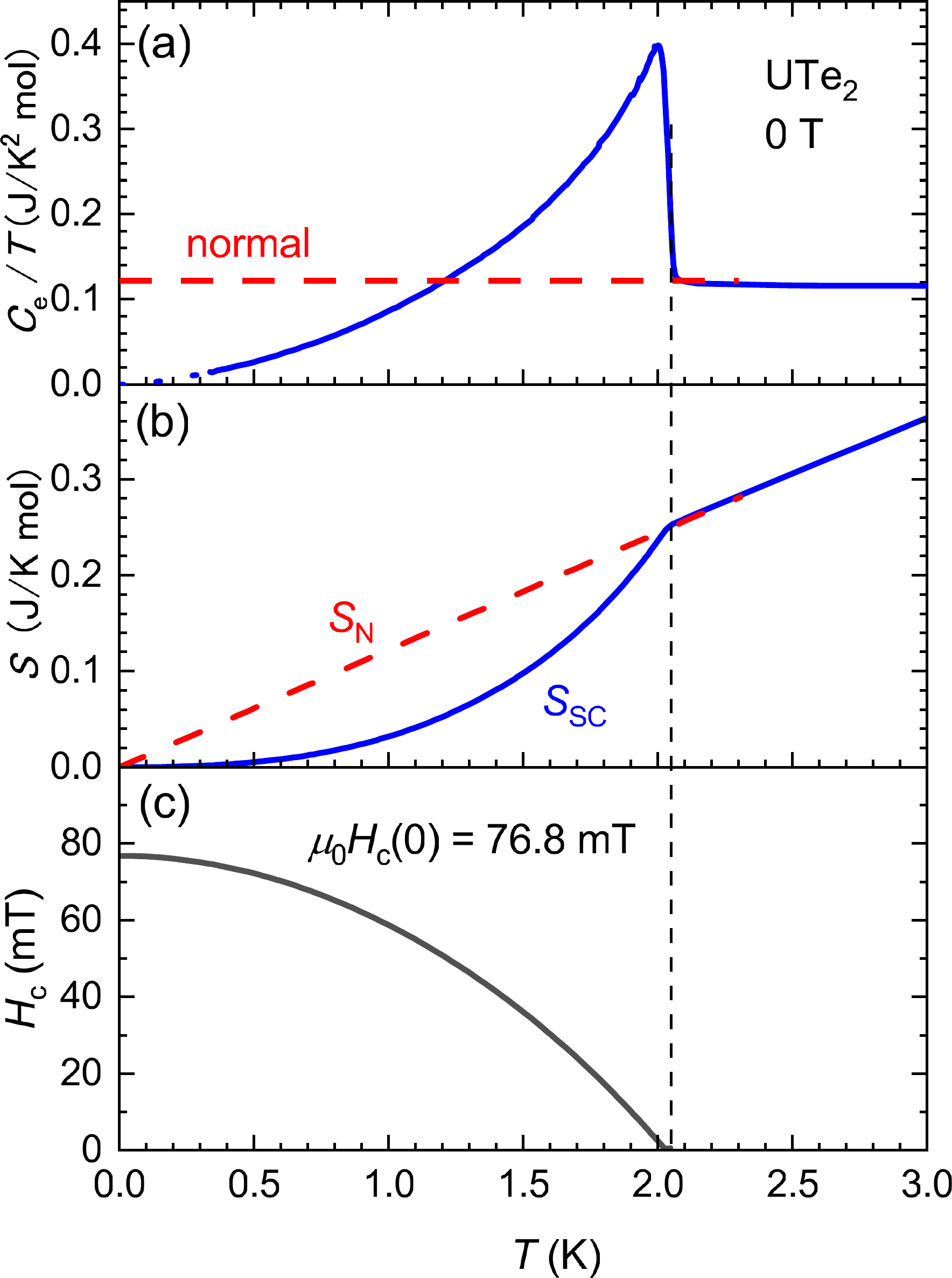}
\end{center}
\caption{(a) Electronic specific heat $C_{\rm e} / T$ of UTe$_2$ as function of temperature. $C_{\rm e}$ was estimated from the subtraction of $C_{\rm ph} \propto T^3$. The dashed line is the expected specific-heat behavior in the normal state. (b) Temperature dependence of the entropy in the normal and SC state. (c) Temperature dependence of the thermodynamic critical field $H_{\rm c}$ deduced from the difference in the free energy $\Delta F$ = $F_{\rm N}$ -$F_{\rm SC}$ = $\int_{0}^{T_{\rm c}}{\Delta S(T) dt}$ = $H_{\rm c}^2/2\mu_0$ of the normal state and the SC states. }
\label{fs2}
\end{figure}

The specific heat of the 2.1~K sample was measured down to 0.3~K in zero field. 
The electronic term of the specific heat $C_{\rm e}/T$, shown in Fig.~1(b) and Fig.~S~\ref{fs2}(a), was determined by the subtracting a phonon contribution ($C_{\rm ph} \propto T^3$). 
$C_{\rm e} / T$ below 0.3~K is the extrapolation from the $T$ dependence between 1 and 0.3~K.
The dashed line is the expected normal state specific-heat behavior, which is assumed to be the $T$ independent $C_{\rm e} / T$ of the 2.1~K value. 
Figure~S~\ref{fs2}(b) shows the temperature dependence of the entropy in the normal and superconducting states. 
The nearly identical entropy values at $T_{\rm c}$ between $S_{\rm N}$ and $S_{\rm SC}$ indicate the satisfaction of the entropy balance in the SC state.
The thermodynamic critical field $H_{\rm c}$ is estimated from the difference of the free energy between the normal and SC states, 
$\Delta F$ = $F_{\rm N}$ -$F_{\rm SC}$ = $\int_{0}^{T_{\rm c}}{\Delta S(T) dt}$ = $H_{\rm c}^2/2\mu_0$. 
From the analysis, the isotropic thermodynamic critical field was estimated as $\mu_0 H_{\rm c}(0)$ = 76.8~mT.

\section{Angle dependence of the $^{125}$Te-NMR spectra in the $bc$ plane}
\begin{figure}[h]
\begin{center}
\includegraphics[width=8cm]{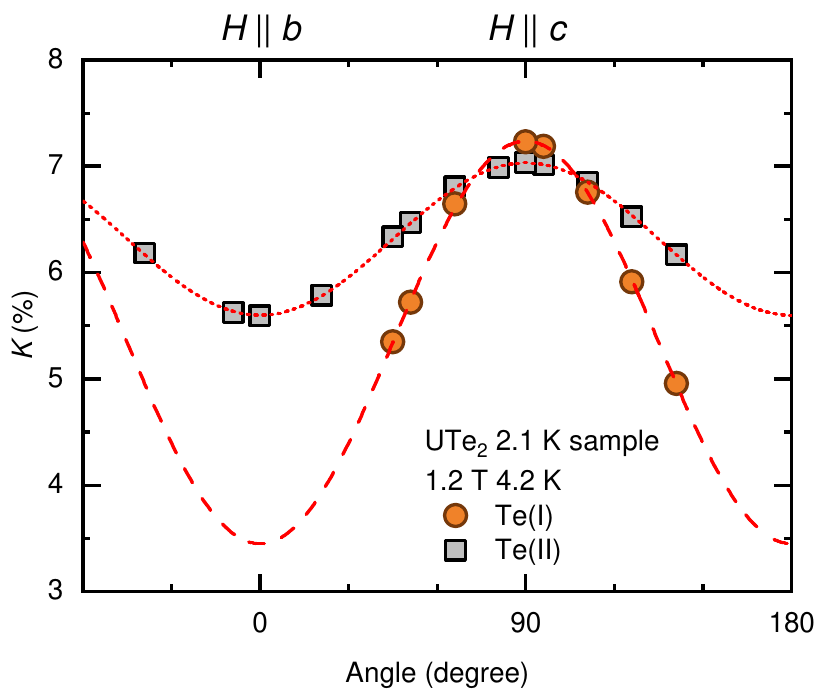}
\end{center}
\caption{The angular dependence of $^{125}$Te-NMR Knight shift at both the Te(I) and Te(II) sites at 4.2~K.
The magnetic field is 1.2 T in the $bc$ plane.
The dashed line represents the Te(I) signal and the dotted line represents the Te(II) signal.}
\label{fs1}
\end{figure}

To apply the magnetic field parallel to each axis, we rotated the single-crystal sample in the $ab$ and $bc$ planes and measured the angle dependence of the Knight shift.  
When $H$ is around the $c$ axis in the $bc$ plane, Te(I) and Te(II) signals are overlapped as shown in Fig.~S~\ref{fs1}, but the two-peak structure remains in $H \parallel c$, which is in contrast to the case of the 1.6~K sample where two peaks become one broad peak\cite{NakaminePRB2021_sm}.
We determined the $c$ axis as the angle at which the Knight shift is maximum.

\section{Superconducting diamagnetic shielding effect in the Knight shift}
In the SC state, the Knight shift decreases owing to the SC diamagnetic shielding effect, and the value of $K_{{\rm dia}, i}$ at the lowest temperature is approximately expressed as\cite{deGennes} 
\begin{equation} \label{eq1}
K_{{\rm dia}, i} = - \frac{H_{{\rm c1}, i}}{H} \frac{\ln\left(\frac{\beta \lambda_d}{\sqrt{2.7\xi_j\xi_k}}\right)}{\ln{\sqrt{\kappa_j\kappa_k}}}.
\end{equation}
Here, $\xi_i$ is the Ginzburg-Landau (GL) coherence length along the $i$ axis with $i, j, k = \{ a, b, c \}$; $\beta$ is a factor that depends on the vortex structure and is 0.38 for the triangular vortex lattice; $\lambda_d$ is the distance between the vortices and is calculated using the relation $\phi_0 = \frac{\sqrt{3}}{2} \lambda_d^2(\mu_0H_{\rm ext})$; and $\kappa$ is the GL parameter.
We estimated $\xi_i$ and $\kappa_i$ with the experimental results of the thermodynamic critical field $\mu_0 H_{\rm c}$ = 76.8 mT estimated from the specific-heat result, and the SC upper critical field along each axis $H_{\rm c2}$ listed in Table \ref{t2}. $H_{\rm c2}$ in $H \parallel b$ is adopted as the upper critical field of the Low-field superconducting phase\cite{Sakai_condmat2022}. Here we used the relation of $H_{{\rm c2}, i} = \phi_0/(2\pi \xi_j \xi_k)$ with the flux quantum $\phi_0$. $H_{\rm c1}$ was estimated from the relation of $H_{{\rm c1}, i} = (H_{\rm c}^2/H_{{\rm c2},i})[\ln(\sqrt{\kappa_j \kappa_k})+0.49]$\cite{PaulsenPRB2021}.    
\begin{table}[htb]
\begin{center}
\caption{\label{t2}Superconducting parameters used for the estimation of $K_{\rm dia}$ at low temperatures. $H_{\rm ext}$ is the external field used for the Knight-shift measurement. }  
\vspace{5mm}
  \begin{tabular}{ccccccc}\hline \hline 
      & & $H \parallel a$ & & $H \parallel b$ & & $H \parallel c$ \\ \hline   
$\mu_0 H_{\rm c2}$ (T) & & 12\cite{Tokiwa_condmat2022}  & & 22\cite{Sakai_condmat2022} & & 17\cite{Tokiwa_condmat2022} \\
$\kappa_i$ & & 293 & & 87.1 & &146\\
$\xi_i$ (nm) & & 3.19 & & 5.85 & & 4.52 \\
$\mu_0 H_{\rm c1}$ (mT) & & 2.56 & & 1.56 & & 1.93 \\   
$\mu_0 H_{\rm ext}$ (T) & & 0.66 & & 0.66 & & 1.2 \\   
$K_{\rm dia} (\%)$ & & 0.08 & & 0.06 & &0.03 \\ \hline \hline
  \end{tabular}
  \end{center}
\end{table}
The evaluated $K_{{\rm dia}, i}$ was subtracted from the estimated reduction of the Knight shift at $T = 0$. 
Here, the demagnetization effect is negligibly small.

\section{$1/T_1T$ in $H \parallel b$ just below $T_{\rm c}$}
\begin{figure}
\begin{center}
\includegraphics[width=7.5cm]{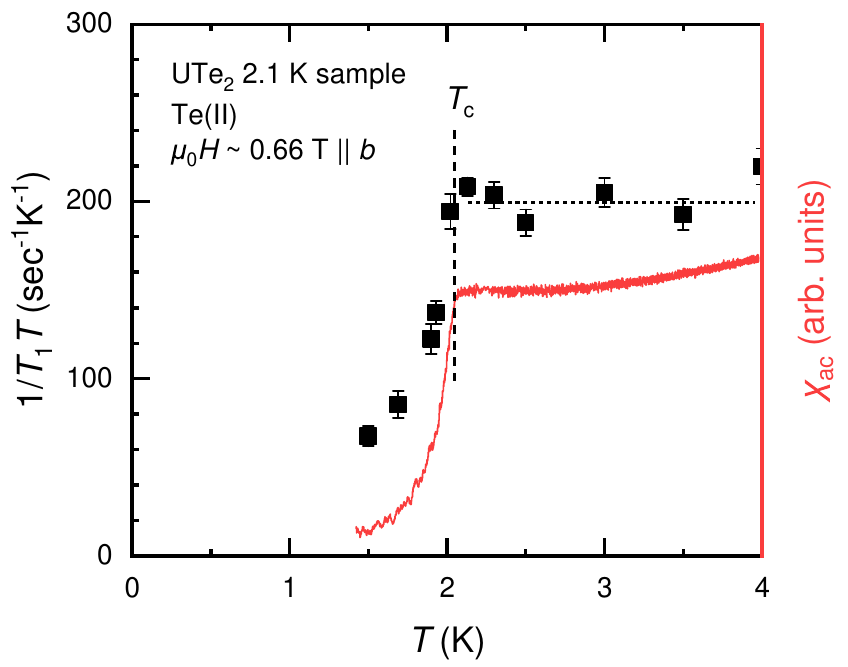}
\end{center}
\caption{The temperature dependence of $1/T_1T$ and $\chi_{\rm ac}$ measured at 
0.66 T which is the same field as the field of the Knight-shift measurement. }
\label{fs3}
\end{figure}

$1/T_1$ was measured at the Te(II) peak in $H \parallel b$ down to 1.5~K.
Figure~S~\ref{fs3} shows the temperature dependence of $1/T_1T$ and $\chi_{\rm ac}$ measured at 0.66 T which is the same field as the field of the Knight-shift measurement.
$1/T_1T$ immediately decreases just below $T_{\rm c}$, showing the absence of the coherence peak.
This excludes the possibility of the conventional $s$-wave superconductivity in UTe$_2$. 

\end{document}